\documentclass[aps,pra,onecolumn,groupedaddress,showpacs]{revtex4}

\usepackage{graphics}
\usepackage{epsfig}
\usepackage{amsmath}

\begin{document}

\title{Deterministic Plug-and-Play for Quantum Communication}

\author{Marco Lucamarini$^{1}$}
\email{marco.lucamarini@roma1.infn.it}

\author{Giovanni Di Giuseppe$^{2}$}



\affiliation{\medskip $^{1}$Dipartimento di Fisica, Universit\`{a}
di Roma ``La Sapienza", I-00185 Roma, Italy. \\
$^{2}$Dipartimento di Fisica, Universit\`{a} di Camerino, I-62032 Camerino (MC), Italy.}


\begin{abstract}
We present a scheme for secure deterministic quantum communication without using entanglement, in a Plug-and-Play fashion.
The protocol is completely deterministic, both in the encoding procedure and in the control one, thus doubling the
communication rate with respect to other setups; moreover, deterministic nature of transmission, apart from rendering
unnecessary bases revelation on the public channel, allows the realization of protocols like `direct communication' and
`quantum dialogue'. The encoding exploits the phase degree of freedom of a photon, thus paving the way to an optical fiber
implementation, feasible with present day technology.
\end{abstract}



\maketitle

\section{INTRODUCTION}

Quantum Key Distribution (QKD) has been the first practical realization of communication based on the laws of the quantum
mechanics. As Bennett and Brassard showed with their pioneer protocol BB84~\cite{bb}, nonorthogonal quantum states, together
with an unjammable classical channel, can be used to prevent an eavesdropper (Eve) from gaining information on the key
without being revealed. There are already different working realizations of BB84~\cite{gis}. Some of them~\cite{Ent-Pol}
exploit entanglement in polarization to encode a random string of bits in two unbiased polarization bases of a single photon,
thus following the original proposal. However, polarization is not the most suitable degree of freedom for fiber
communication because birefringence usually fluctuates randomly in a fiber, making the encoded information impossible to
decode. The most appealing setup for real telecom communications is based on the phase degree of freedom.

\begin{figure}[h]
\begin{center}
\includegraphics[width=.75\textwidth]{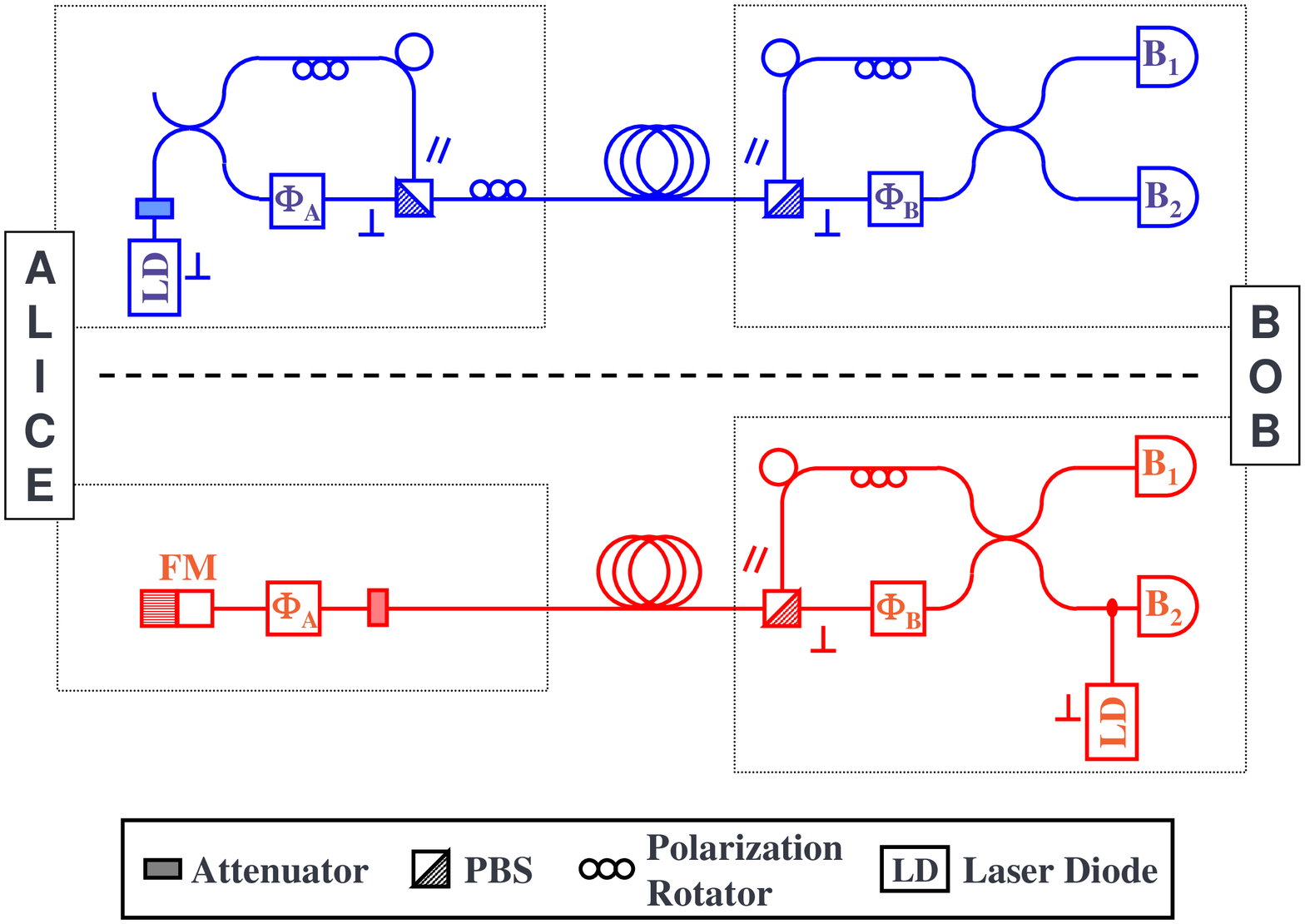}
\end{center}
\caption{\label{BB84} Schematics of two interferometers for BB84.
The upper one, in blue, has been used for QKD up to 30 km
    ; the lower one, in red, is the Plug-and-Play setup used for QKD
    up to 67 km.}
\end{figure}

As an example we consider the setup shown in the upper part of
Fig.\ref{BB84}, that has been used for a QKD up to 30
km~\cite{mar}. In this arrangement a relative phase
$\Phi_{A}\equiv\{0,\pi/2,\pi,3\pi/2\}$ between two time-bins of a
photon state is realized by the sender (Alice) with an unbalanced
interferometer and a phase modulator. The receiver (Bob) measures
incoming bins by means of a second interferometer, matched to the
sender's one, set to $\Phi_{B}\equiv\{0,\pi/2\}$. With a
probability of 50\% the phase difference between Alice and Bob's
interferometers will be $0$ or $\pi$, and their measures will be
correlated; in the other half of cases measures will be discarded
by the two users.

Another realization of phase encoded QKD, namely the
``Plug-and-Play" setup~\cite{gis}, is reported in the lower part
of Fig.\ref{BB84}, and it has been used for QKD up to a distance
of 67 km~\cite{stu}. Using an intense pulse, Bob populates two
time-bins with relative phase $\Phi_{B}=0$, and sends them to
Alice. Alice applies a relative phase
$\Phi_{A}\equiv\{0,\pi/2,\pi,3\pi/2\}$, reflects the pulses on a
Faraday mirror, let them pass through an attenuator, and sends
them back to Bob. The two pulses retrace in their backward path
all the fluctuations they suffered in the forward path, thus
arriving at Bob's interferometer ready for being
interferometrically revealed. As in the previous scheme Bob sets
at random $\Phi_{B}\equiv\{0,\pi/2\}$, and measures the outcomes,
completing the \textit{probabilistic} QKD run pertaining to BB84.
This mechanism of automatic compensation of noise earned this
setup the name of ``Plug-and-Play".

Recently Bostr\"{o}m and Felbinger~\cite{bf} presented a variant
for QKD, and named it ``Ping-Pong" cryptography (PP) for its
peculiarity of a forward and backward use of the quantum channel.
This peculiarity seems to suit perfectly with the forward and
backward dynamics typical of Plug-and-Play setups, as the one
described above, and this motivated the present work. The main
advantage of PP is the deterministic nature of its
\textit{encoding-decoding} procedure; its main flow is that it has
been proved to be not completely secure~\cite{cai,woj}. A number
of variants, still based on the polarization degree of freedom of
a photon, have been proposed to solve this problem, some making
use of entanglement~\cite{PP-modified}, and others making no use
of it~\cite{PP_mod_no_ent,luc}.

We envisage here a protocol with phase-encoded information to have a secure, completely deterministic, quantum communication.
A few variants and applications of the protocol are also discussed.

\section{PROTOCOL}

The receiving user, Bob, prepares a photon in a superposition of
two time-bins with a relative phase $\Phi_{B}$ randomly chosen
between the values $\{0,\pi/2,\pi,3\pi/2\}$. He sends the photon
to Alice, the transmitting user, who chooses one of two possible
tasks, {\it control mode} (CM) or {\it message mode} (MM), with
probability $c$ and $1-c$ respectively; the former realizes a
control on the security of the channel, the latter the
deterministic communication between the users.

\underline{MM:} Alice encodes a bit of information with an unitary
operation on the two time-bins: she can either apply the identity
operation, choosing $\Phi_{A}=0$ and encoding a `0', or introduce
a `phase-flip', by applying a phase $\Phi_{A}=\pi$ between the
time-bins and encoding a `1'. The photon is sent back to Bob who
measures it with his apparatus set with the same phase $\Phi_{B}$
he prepared initially. In this way Bob's measurement is {\it
deterministic}, because the initial phase of the state (that he
does know) is changed by an amount of 0 or $\pi$; he can thus
guess Alice's operation without any needs of a classical channel.

\underline{CM:} Alice detects the photon with her interferometer
randomly set to $\Phi_{A}=0$ or $\Phi_{A}=\pi/2$. After that Alice
prepares a {\it new photon state} with a phase
$\Phi'_{A}=\Phi_{A}+\pi/2$, and sends it back to Bob who,
analogously to what he did in MM, measures it with the same phase
$\Phi_{B}$ he prepared initially. We notice that if phase
difference between Bob's interferometer and Alice's one is $0$ or
$\pi$ then the two users share correlated information, while in
the other two cases they do not. It is straightforward to realize
that Alice procedure in preparing the new photon let the two
users' interferometers to be necessarily correlated either in the
forward path or in the backward one. This entails that the CM of
our protocol results as deterministic as the MM is, thus achieving
that doubling of the whole rate transmission we mentioned before.
None of the qubits, neither destined to MM nor to CM, is
discarded.

\section{SECURITY AND IMPLEMENTATION}

Given un unjammable public channel two users can exchange
information in a secure manner by means of the above protocol. Eve
can try to gain information by inquiring the phase of the photon
both on the forward and in the backward path; alternatively, she
can prevent Bob from gaining information (DoS attack~\cite{cai})
by randomly measuring the state of the traveling photon; finally,
on a lossy channel, she can conceal her presence behind losses
~\cite{woj}. Nevertheless no attack can remain undetected by the
control procedure since it is akin to a BB84 check test, performed
either on the forward path or on the backward one. This guarantees
the protocol is unconditionally secure, as BB84 is, even if the
security threshold may be different. Our protocol gives a
probability of 25\% to detect DoS attack as well as a general
attack providing Eve with full information.

\begin{figure}[h]
\begin{center}
\includegraphics[width=.75\textwidth]{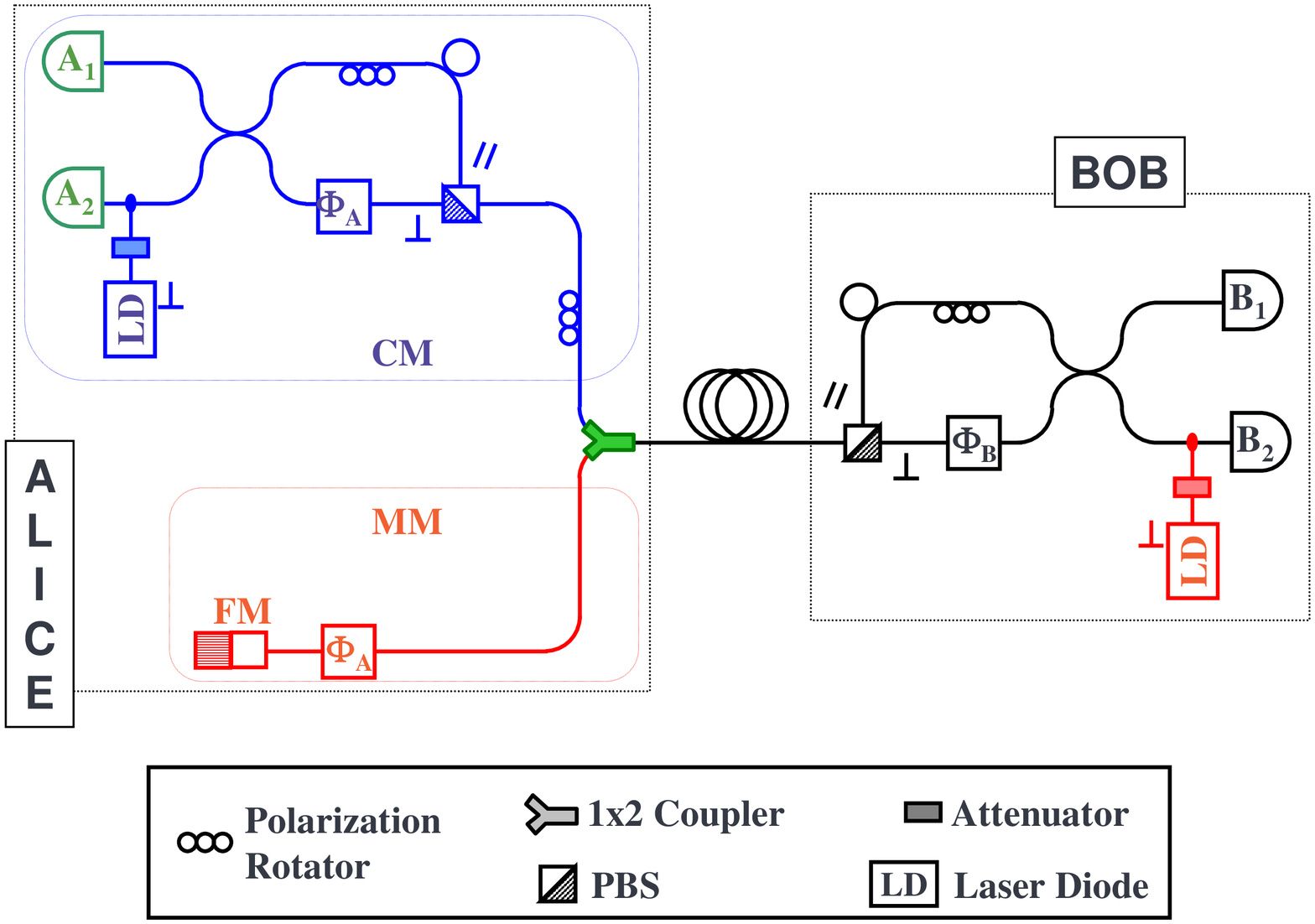}
\end{center}
\caption{\label{PPPP84} Final setup for deterministic
Plug-and-Play cryptography. It merges the two interferometers of
Fig.1, reported in blue (CM) and red (MM). The black represents
the common elements between the two, while the green indicates new
components. An automatic switch between MM and CM is performed by
a 1x2 coupler. }
\end{figure}

The practical implementation requires to gather the two
interferometers of Fig.\ref{BB84}. The resulting scheme is shown
in Fig.\ref{PPPP84}: the upper interferometer is devoted to CM,
the lower to MM, and they are connected by a 1x2 fiber-coupler.
The signal prepared by Bob goes at random into CM or into MM with
probability $c$ and $1-c$, respectively. Later on, when the photon
is on the backward path, this device redirects it to Bob.
The MM-interferometer resembles the one used for testing BB84 up
to 67 km~\cite{stu}. It is a typical Plug-and-Play scheme, with
the unique difference that the attenuator is inserted just after
the laser-diode (LD), at Bob's side. A laser pulse with a mean
photon number much lower than unity goes through Bob's
interferometer and comes out into two time-bins with a relative
phase $\Phi_{B}$ and opposite polarizations. In the Plug-and-Play
setup an intense pulse is sent to Alice to provide a trigger
signal for Alice's electronics. In our case, as the pulse is
already attenuated at Bob's side, a further synchronization
system, such as usually used in other practical QKD
setup~\cite{gis,Ent-Pol,mar}, is necessary. After reflection on
the Faraday mirror, the photon travels back to Bob, retracing
exactly those paths that let him compensate any undesired phase:
this makes unnecessary the stabilization of the polarization.
Eventually, Raileigh backscattering is not a problem in this case
because of the low value of the average photon number from the
beginning of the protocol.

The CM-interferometer resembles the one used for BB84 QKD up to a
distance of 30 km~\cite{mar}. As a time-polarization division
technique is implemented, this task is sensible to random phase
changes, and needs adjustments during the protocol runs. A value
of $\sim 0.6$~ rad/min for slow thermal drift in the
interferometer was estimated, thus requiring a compensation step
every $\sim 5s$~\cite{mar}. Also in this case Alice must
synchronize her electronics with detectors, phase-modulator, and
laser-diode using a suitable synchronization system.

\section{CONCLUSIONS}

We proposed a setup that fully exploits the potentialities of the
two-way unbalanced interferometers of certain implementations of
BB84 to merge the features of Plug-and-Play setup for BB84 and
Ping-Pong cryptography. The scheme is completely deterministic,
both in the encoding-decoding procedure and in the control one,
thus achieving a doubling in the rate of information transmission.
It seems to be quite practical to implement since it is based on
existing working implementations.

Besides, due to its deterministic nature, several other advantages
are available. Bases revelation on the public channel is avoided,
and this feature, together with a slightly different Alice's
preparation procedure during the CM, leads to a QKD more secure
than that achieved by means of BB84~\cite{luc}. Furthermore a
`quantum direct communication' (QDC) between users is also
possible~\cite{bf,luc}. Finally, we notice that once QDC is
available also the novel protocol called `quantum
dialogue'~\cite{ngu} can be achieved: if Alice delivered a message
to Bob by means of a secure QDC, then Bob can use the message just
received as the key of encryption of his answer, and simply
communicate it to Alice on the public channel.




\end{document}